\RequirePackage{arydshln}
\documentclass[aps,twocolumn,nofootinbib,superscriptaddress,preprintnumbers,pra,10pt,floatfix]{revtex4-1}

\usepackage[utf8]{inputenc}

\usepackage{float}
\usepackage{amsmath,amssymb}
\usepackage{dsfont} 
\usepackage{hyperref}
\usepackage{graphicx}
\usepackage{enumitem}
\usepackage{mathtools}
\usepackage{bbold}
\usepackage{multirow}
\usepackage{ytableau}
\usepackage{youngtab}
\usepackage{braket}
\usepackage{soul}

\usepackage{colortbl} 
\definecolor{Gray}{gray}{0.95}
\definecolor{RGray}{gray}{0.90}
\definecolor{CGray}{gray}{0.92}

\usepackage{arydshln}

\makeatletter
\g@addto@macro\bfseries{\boldmath}
\makeatother

\makeatletter
\renewcommand\paragraph{\@startsection{paragraph}{4}{\z@}%
                                    {3.25ex \@plus1ex \@minus.2ex}%
                                    {-1em}%
                                    {\normalfont\normalsize\bfseries}}
\makeatother

\allowdisplaybreaks
\begin{document}

\preprint{ZU-TH-08/20}
\preprint{LAPTH-014/20}

\title{Third-family quark-lepton unification with a fundamental composite Higgs}

\author{Javier Fuentes-Mart\'{\i}n}
\email{fuentes@physik.uzh.ch}
\affiliation{Physik-Institut, Universit\"at Z\"urich, CH-8057 Z\"urich, Switzerland}
\author{Peter Stangl}
\email{peter.stangl@lapth.cnrs.fr}
\affiliation{Laboratoire d'Annecy-le-Vieux de Physique Th\'eorique, 
UMR5108, CNRS, Universit\'e de Savoie Mont-Blanc,
B.P. 110, F-74941 Annecy-le-Vieux Cedex, France\looseness=-1}

\begin{abstract}
\vspace{5mm}
We present a model for third-family quark-lepton unification at the TeV scale featuring a composite Higgs sector. The model is based on a variant of the Pati-Salam model, the so-called 4321 model, consisting of the gauge group $SU(4)\times SU(3)^\prime\times SU(2)_L\times U(1)_X$. The spontaneous symmetry breaking to the SM gauge group is triggered dynamically by a QCD-like confining sector. The same strong dynamics also produces the Higgs as a pseudo Nambu-Goldstone boson, connecting the energy scales of both sectors.  The model predicts a massive $U_1$ vector leptoquark coupled dominantly to the third generation, recently put forward as a possible solution to the $B$-meson anomalies.
\vspace{3mm}
\end{abstract}

\maketitle

\allowdisplaybreaks

\section{Introduction}\label{sec:intro}

The LHC discovery of a $125~\mathrm{GeV}$ Higgs boson provides the Standard Model (SM) with its final eluding piece that completes it. However, it is still unclear whether the Higgs represents the first indication of a yet unknown \emph{natural} theory, or just an ingredient of an \emph{unnatural} Higgs sector. While nowadays some degree of tuning seems unavoidable, one viable solution to the naturalness problem is that of the Higgs being a composite particle arising from a strongly-coupled sector.

Historically, the main hurdles for composite scenarios have been electroweak (EW) precision data and flavor violation. In models where the Higgs is a pseudo Nambu-Goldstone boson (pNGB), it is possible to separate EW and compositeness scales~\cite{Kaplan:1983fs}. Corrections to the EW precision parameters are then under control if the pNGB decay constant is around (or above) the TeV~\cite{Contino:2010rs}. The flavor problem is commonly solved by making the SM fermions partially composite~\cite{Kaplan:1991dc}. This way, one can generate the required Yukawa interactions with the Higgs, while having partial protection against flavor violating observables. Even in this case, a non-trivial flavor structure is necessary to keep the pNGB decay constant around the TeV~\cite{Redi:2011zi,Redi:2012uj,Niehoff:2015iaa}. An alternative approach consists in generating the Yukawas via bilinear terms of the form $\bar\psi_{\rm SM}\,\mathcal{H}\,\psi_{\rm SM}$, with $\mathcal{H}\equiv\bar\Psi\Psi$ being a composite operator~\cite{Dimopoulos:1979es,Eichten:1979ah}. This solution is often disregarded based on the argument that the dynamics generating these operators is also likely to produce large flavor violation from operators of the form $(\bar\psi_{\rm SM}\,\psi_{\rm SM})^2$, which is strongly constrained experimentally. Such a conclusion drastically changes if one invokes flavor symmetries. Indeed, assuming that the underlying dynamics respects an approximate $U(2)^5$ flavor symmetry~\cite{Barbieri:2011ci,Blankenburg:2012nx,Barbieri:2012uh}, it is possible to generate the required Yukawa couplings, without conflicting any flavor bound on the four-fermion operators.

While there is no direct signal of New Physics (NP) at the end of 
LHC Run II, present $B$-physics data show intriguing hints of lepton flavor universality violation that the SM cannot explain, the so-called $B$ anomalies~\cite{Aaij:2013qta,Aaij:2014ora,Aaij:2015oid,Aaij:2017vbb,Aaij:2019wad,Lees:2012xj,Huschle:2015rga,Aaij:2015yra,Aaij:2017tyk,Aaij:2017uff,Abdesselam:2019dgh}. Although the statistical significance of each anomaly is well below the discovery level, the overall set of deviations is very consistent, and a coherent NP picture seems to be emerging~\cite{Aebischer:2019mlg,Alguero:2019ptt,Crivellin:2018yvo}. The NP scale inferred from these anomalies is a few TeV, sustaining the hope that such NP sector might be related to the solution of the hierarchy problem~\cite{Gripaios:2014tna,Niehoff:2015bfa,Barbieri:2016las,Megias:2016bde,Barbieri:2017tuq,Sannino:2017utc,Marzocca:2018wcf}. Moreover, the non-trivial flavor structure suggested by the data is consistent with the previously mentioned $U(2)^5$ flavor symmetry~\cite{Fuentes-Martin:2019mun,Buttazzo:2017ixm,Barbieri:2015yvd}, pointing to a possible solution to the flavor problem in composite models~\cite{Barbieri:2016las,Barbieri:2017tuq,Marzocca:2018wcf} and (or) the SM flavor puzzle~\cite{Bordone:2017bld,Greljo:2018tuh,Barbieri:2019zdz}.

The $B$ anomalies have triggered a renewed interest in models of low-scale quark-lepton unification. Indeed, one of the most popular explanations involves the $U_1$ vector leptoquark~\cite{Barbieri:2015yvd,Alonso:2015sja,Buttazzo:2017ixm,Calibbi:2017qbu}, transforming under the SM gauge group as $(\mathbf{3},\mathbf{1},2/3)$. Interestingly, this is the same leptoquark appearing in the Pati-Salam model~\cite{Pati:1974yy}. However, the Pati-Salam leptoquark cannot accommodate this data, since it has to be very heavy to satisfy the tight bounds derived from its couplings to light SM fermions. The search for a renormalizable model with a TeV-scale $U_1$ leptoquark has led to the so-called \emph{4321 models}~\cite{Georgi:2016xhm,Diaz:2017lit,DiLuzio:2017vat,Blanke:2018sro,DiLuzio:2018zxy,Bordone:2017bld,Greljo:2018tuh,Bordone:2018nbg,Cornella:2019hct}, based on the gauge group $\mathcal{G}_{4321}\equiv SU(4)\times SU(3)^\prime\times SU(2)_L\times U(1)_X$. These models present several theoretically appealing features that go beyond the explanation of the $B$ anomalies. For instance, they bring the possibility of unifying third-generation quarks and leptons at energy scales as low as TeV, introducing an (approximate) accidental $U(2)^5$ flavor symmetry, and can naturally explain the smallness of the Cabibbo-Kobayashi-Maskawa (CKM) mixing with the third family.

\begin{figure*}[t]
    \centering
    \includegraphics[width=0.8\textwidth]{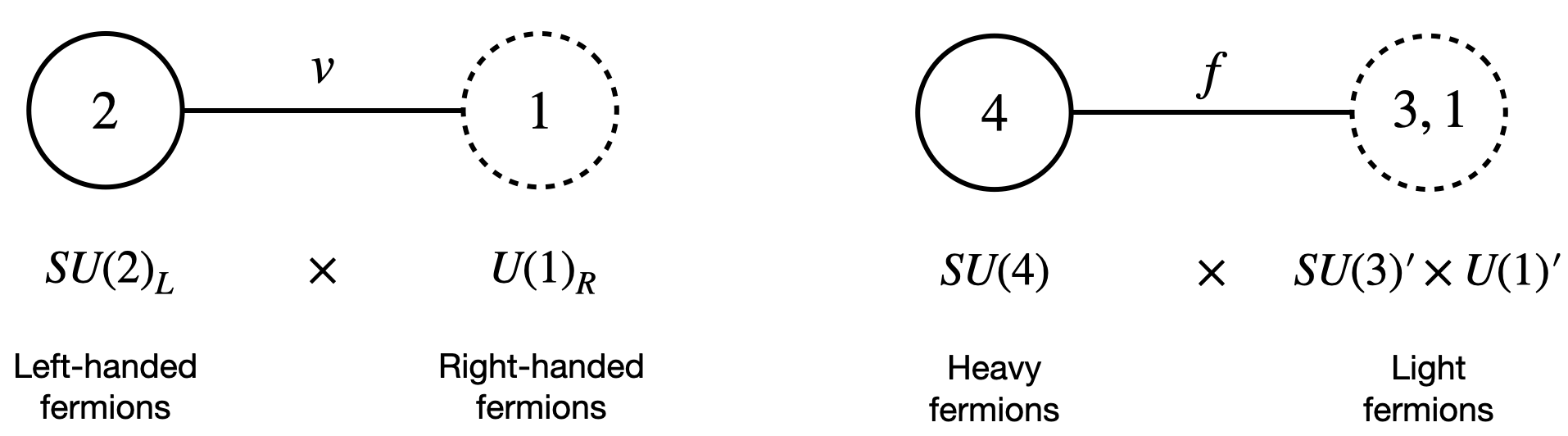}
    \caption{Moose diagrams for the EW sector (left) and the 43(2)1 model (right). Following the notation in~\cite{Chivukula:2003wj}, we draw a solid circle when the entire global symmetry is gauged, and a dashed circle when a subgroup is. The solid lines represent sigma models that break the symmetries to which they are attached down to the diagonal subgroups. The $U(1)_X$ gauge factor in 4321 models is the diagonal combination of $U(1)_R$ and $U(1)^\prime$ symmetries.}\label{fig:moose}
\end{figure*}

\begin{figure*}[t]
    \centering
    \includegraphics[width=\textwidth]{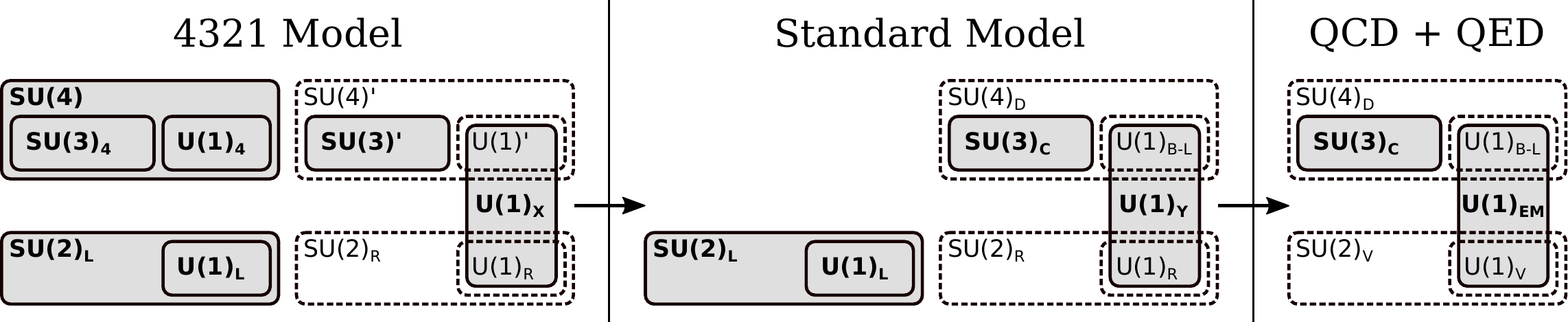}
    \caption{Group structure and symmetry breaking pattern in 4321 models. Dashed boxes correspond to partially gauged global symmetries, while bold names correspond to gauge symmetries. The 4321 breaking pattern $SU(4)\times SU(4)'\to SU(4)_D$ leads to the subgroup breakings $SU(3)_4\times SU(3)'\to SU(3)_c$ and $U(1)_4\times U(1)'\to U(1)_{B-L}$. The gauged $U(1)$ symmetries break as $U(1)_4\times U(1)_X\to U(1)_{Y}$. The SM breaking pattern $SU(2)_L\times SU(2)_R\to SU(2)_V$ leads to the subgroup breaking $U(1)_L\times U(1)_R\to U(1)_{V}$. The gauged $U(1)$ symmetries break as $U(1)_L\times U(1)_Y\to U(1)_{\rm EM}$.}
    \label{fig:4321breaking}
\end{figure*}

It is useful to compare the 4321 symmetry breaking pattern to that of the EW sector, given the astonishing similarity between the two. In the limit of vanishing gauge and Yukawa couplings, the SM Higgs sector has an $SU(2)_L \times SU(2)_R$ global symmetry. The Higgs vacuum expectation value (vev) spontaneously breaks this global symmetry to the diagonal $SU(2)_V$ subgroup. The three resulting NGBs become would-be NGB due to the partial gauging of the global symmetry. In the SM, all three generators of $SU(2)_L$ and the diagonal generator $T_R^3$ of $SU(2)_R$ are gauged (see Figure~\ref{fig:moose}).\footnote{Actually, what is gauged in the SM is the linear combination $Y=T_R^3+\tfrac{1}{2}X_{B-L}$, where $Y$ is the hypercharge generator and $X_{B-L}$ is the generator of the baryon minus lepton number symmetry, cf. Figure~\ref{fig:4321breaking}}. Hence, the global symmetry breaking leads to the breaking of the EW gauge group, $SU(2)_L\times U(1)_Y$, down to the diagonal $U(1)_{\rm em}$ electromagnetic subgroup, and the three NGBs become the longitudinal polarizations of the three massive vector bosons $W_\mu^\pm$ and $Z_\mu$.

Likewise, in the limit of vanishing gauge and Yukawa couplings, 4321 models have an additional $SU(4)\times SU(4)^\prime$ global symmetry that is spontaneously broken to the diagonal $SU(4)_D$ by the vev of a bi-fundamental scalar, producing 15 NGBs. Also in this case, the global symmetry is partially gauged. More precisely, the full $SU(4)$ group and the $SU(3)^\prime\times U(1)^\prime$ subgroup of $SU(4)^\prime$ is gauged (see Figure~\ref{fig:moose}). The global symmetry breaking leads to the breaking of the $SU(4)\times SU(3)^\prime\times U(1)^\prime$ gauge group to its diagonal subgroup $SU(3)_D\times U(1)_D$, which is identified with QCD times (part of) hypercharge. As a result, all 15 NGBs become the longitudinal polarizations of massive vector bosons: the coloron (a hypercharge neutral octet of $SU(3)_c$), the $U_1$ leptoquark, and the SM neutral $Z'$. The coloron and the $Z'$ have the gluon and hypercharge gauge bosons as massless partners. In this regard, they are analogous to the SM $Z$, which has the photon as a massless partner. The leptoquark transforms in the (anti-)fundamental of the unbroken gauge group and does not have a massless partner. It is thus analogous to the SM $W$, which is charged under the unbroken electromagnetic gauge group and does not have a massless partner either.

As it is well known, QCD with $N_f$ quark flavors has an $SU(N_f)_L \times SU(N_f)_R$ global symmetry that is spontaneously broken to its diagonal $SU(N_f)_V$ subgroup. In this case, the breaking is not induced by the vev of a scalar field, but by the quark condensate that forms after QCD becomes strongly coupled. While the scale of this breaking is far too low to explain the observed $W$ and $Z$ masses, it inspired the idea that a scaled-up version of QCD, known as technicolor~\cite{Weinberg:1975gm,Susskind:1978ms,Farhi:1980xs}, could be responsible for EW symmetry breaking. After the discovery of the Higgs boson, traditional technicolor was excluded. However, a technicolor-like breaking is still possible for the 4321 symmetry.

Given the apparent coincidence of scales between composite Higgs models and the $B$ anomalies, and the fact that both seem to benefit from the same underlying flavor symmetries, we entertain the possibility of having both 4321 and EW symmetries broken by the same strongly-coupled ``hypercolor" (HC) group. Our construction resembles a generalization of technicolor for the  4321 symmetry breaking, while the EW symmetry is broken by the vev of a composite Higgs arising as a pNGB of the same strong dynamics. Since we provide a description of the fundamental HC Lagrangian, such a Higgs is usually referred to as ``fundamental composite Higgs"~\cite{Cacciapaglia:2014uja} to distinguish it from other constructions, like the holographic composite Higgs~\cite{Contino:2003ve}.

The outline of this letter is as follows: In Section~\ref{sec:4321}, we introduce the 4321 models and define our conventions. The idea of a technicolor-like breaking of the 4321 symmetry is developed in Section~\ref{sec:TC4321}, while Section~\ref{sec:Higgs} is devoted to the discussion of the composite Higgs sector. We conclude in Section~\ref{sec:Conclusions}.

\section{The 4321 model(s)}\label{sec:4321}

\subsection{Gauge sector}
The 4321 models are defined by the gauge group $\mathcal{G}_{4321}\equiv SU(4)\times SU(3)^\prime\times SU(2)_L\times U(1)_X$. We denote the respective gauge fields by $H_\mu^A$, $C_\mu^a$, $W_\mu^I$ and $B_\mu^\prime$, and the gauge couplings by $g_4$, $g_3$, $g_L$ and $g_1$, with indices $A=1,\dots,15$, $a=1,\dots,8$ and $I=1,2,3$. The group structure and symmetry breaking pattern of the 4321 model down to QCD + QED is described in Figure~\ref{fig:4321breaking}. The SM gauge group is embedded in the 4321 gauge group, with $SU(3)_c\times U(1)_Y\equiv \left[SU(4)\times SU(3)^\prime\times U(1)_X\right]_{\rm diag}$ and $SU(2)_L$ corresponding to the SM one. The hypercharge is defined in terms of the $U(1)_X$ charge, $X$, and the $SU(4)$ generator  $T_4^{15}=\frac{1}{2\sqrt{6}}\,\mathrm{diag}(1,1,1,-3)$ by $Y=X+\sqrt{2/3}\,T_4^{15}$. In analogy to the EW sector, it is convenient to define the mixing angles $\theta_{1,3}$, which relate the 4321 gauge couplings to the SM ones,
\begin{align}\label{eq:SMgaugeCrel}
\begin{aligned}
g_c&=g_4\sin\theta_3=g_3\cos\theta_3\,,\\
g_Y&=\sqrt{\tfrac{3}{2}}\,g_4\sin\theta_1=g_1\cos\theta_1\,,
\end{aligned}
\end{align}
with $g_c$ and $g_Y$ denoting the $SU(3)_c$ and $U(1)_Y$ gauge couplings, respectively. These relations imply that, once we fix the SM gauge couplings, there is only one free gauge coupling in the 4321 models, which we choose to be $g_4$. Also note that, since ${g_c>g_Y}$, the relation ${\sin\theta_3>\sin\theta_1}$ holds for any value of $g_4$. 

In terms of the original 4321 gauge bosons, the SM gluon, $G_\mu^a$, and hypercharge gauge boson, $B_\mu$, are given by
\begin{align}
\begin{aligned}
B_\mu&=\sin\theta_1\,H_\mu^{15}+\cos\theta_1\, B^\prime_\mu\,,\\
G_\mu^a&=\sin\theta_3\,H_\mu^a+\cos\theta_3\, C_\mu^a\,.
\end{aligned}
\end{align}
Apart from these, the 4321 gauge sector contains three additional gauge bosons, transforming under the SM subgroup as $U_1\sim(\mathbf{3},\mathbf{1},2/3)$, $Z^\prime\sim(\mathbf{1},\mathbf{1},0)$, and ${G^\prime\sim(\mathbf{8},\mathbf{1},0)}$. After the spontaneous symmetry breaking ${\mathcal{G}_{4321}\to\mathcal{G}_{\rm SM}}$ takes place, these additional gauge bosons
acquire the masses
\begin{align}\label{eq:4321GaugeMasses}
M_U&=\frac{1}{2}\,g_4\,f_U\,,&&& M_{Z^\prime,G^\prime}&=\frac{1}{2}\,\frac{g_4}{\cos\theta_{1,3}}\,f_{Z^\prime,G^\prime}\,,
\end{align}
with the values of $f_{U,Z^\prime, G^\prime}$ depending on the specific vev structure that triggers 4321 symmetry breaking. For instance, in the models in~\cite{DiLuzio:2018zxy,Cornella:2019hct}, this breaking takes place through the vevs of the scalar fields $\Omega_1\sim(\mathbf{\bar 4},\mathbf{1},\mathbf{1},-1/2)$, $\Omega_3\sim(\mathbf{\bar 4},\mathbf{3},\mathbf{1},1/6)$ and $\Omega_{15}\sim(\mathbf{15},\mathbf{1},\mathbf{1},0)$,
\begin{align}\label{eq:4321vevs}
\langle \Omega_1^\alpha\rangle&=\frac{\omega_1}{\sqrt{2}}\,\delta_{\alpha4}\,, & \langle \Omega_3^{\alpha i}\rangle&=\frac{\omega_3}{\sqrt{2}}\,\delta_{\alpha i}\,, &
\langle \Omega_{15}^A\rangle&=
\frac{\omega_{15}}{\sqrt{2}}\,\delta_{A 15}\,,
\end{align}
where $\alpha=1,2,3,4$, $i=1,2,3$, and $\Omega_{15}=\Omega_{15}^A\,T_4^A$, with $T_4^A$ being the $SU(4)$ generators normalized so that $\mathrm{Tr}(T_4^AT_4^B)=1/2\,\delta_{AB}$. This vev structure yields the following values for $f_{U,Z^\prime, G^\prime}$:
\begin{align}
f_U&=\sqrt{\omega_1^2+\omega_3^2+\frac{4}{3}\,\omega_{15}^2}\,,&&&
f_{Z^\prime}&=\sqrt{\frac{3}{2}\,\omega_1^2+\frac{1}{2}\,\omega_3^2}\,,\nonumber\\
f_{G^\prime}&=\sqrt{2}\,\omega_3\,.
\end{align}
Note that these vevs break the global $SU(4)_D$ discussed in Section~\ref{sec:intro}, unless $\omega_1=\omega_3$ and $\omega_{15}=0$. In this case, one has $f_U=f_{Z^\prime}=f_{G^\prime}$. Additional vev structures have been discussed in~\cite{DiLuzio:2018zxy}. As shown in this reference, it is not possible to significantly decouple $f_{Z^\prime}$ from $f_U$, irrespective of the chosen vev structure.

\subsection{Fermion content}

There are several possible embeddings of the SM fields into 4321 representations (see Appendix~\ref{app:SMfermionEmbedding}). While many of the results of this letter are independent of this embedding and apply also to other 4321 models, we focus on the implementation realizing third-family quark-lepton unification~\cite{Bordone:2017bld,Greljo:2018tuh,Cornella:2019hct}. In this implementation, first and second families are charged as in the SM under $SU(3)^\prime\times SU(2)_L\times U(1)_X$, while third-family quarks and leptons are unified in $SU(4)$ multiplets (see Table~\ref{tab:SMfieldcontent}).\footnote{%
Charging light and third generation quarks under two different groups that yield QCD as a diagonal subgroup is reminiscent of topcolor models~\cite{Hill:1991at,Martin:1991uz,Martin:1992aq,Martin:1992mj,Hill:1994hp}.}
Note that we have introduced a third-generation right-handed neutrino to complete the corresponding right-handed $4$-plet.

This model provides an example of third-family quark-lepton unification at an energy scale which can be considerably below the grand unification scale. As in the Pati-Salam model, the proton is stable due to the presence of an accidental global symmetry at the level of renormalizable operators~\cite{DiLuzio:2017vat,Bordone:2017bld}. Moreover, if the Higgs is embedded into a singlet representation of $SU(4)$, the model predicts equal Higgs Yukawa couplings at the unification scale for bottom and tau, and for top and tau-neutrino. This is a good approximation to the observed values of bottom and tau masses. The required (small) mass splitting between the two can be obtained from additional $SU(4)$-breaking sources, as discussed in Section~\ref{subsec:Yukawas}. Naturally light neutrino masses with a low $SU(4)$-breaking scale can be realized through an inverse see-saw mechanism by introducing additional gauge-singlet fermions~\cite{Greljo:2018tuh,Perez:2013osa}.

\begin{table}[t]
\begin{center}
\begin{tabular}{|c|c|c|c|c|}
\hline
Field & $SU(4)$ & $SU(3)'$ & $SU(2)_L$ & $U(1)_X$ \\
\hline
\hline
$\ell^i_L$ & $\mathbf{1}$ & $\mathbf{1}$ & $\mathbf{2}$ & $-1/2$ \\
$e^i_R$ & $\mathbf{1}$ & $\mathbf{1}$ & $\mathbf{1}$ & $-1$ \\ 
$q^i_L$ & $\mathbf{1}$ & $\mathbf{3}$ & $\mathbf{2}$ & $1/6$ \\
$u^i_R$ & $\mathbf{1}$ & $\mathbf{3}$ & $\mathbf{1}$ & $2/3$  \\
$d^i_R$ & $\mathbf{1}$ & $\mathbf{3}$ & $\mathbf{1}$ & $-1/3$  \\
$\psi_L$ & $\mathbf{4}$ & $\mathbf{1}$ & $\mathbf{2}$ & $0$ \\ 
$\psi_R^+$ & $\mathbf{4}$ & $\mathbf{1}$ & $\mathbf{1}$ & $1/2$ \\  
$\psi_R^-$ & $\mathbf{4}$ & $\mathbf{1}$ & $\mathbf{1}$ & $-1/2$ \\[1pt]
\hline
\end{tabular}
\end{center}
\caption{SM fermion content and $4321$ representations. Here $i=1,2$, $\psi_L\equiv(q_L^3\; \ell_L^3)^\intercal$, $\psi_R^+\equiv(u_R^3\; \nu_R^3)^\intercal$, and $\psi_R^-\equiv(d_R^3\; e_R^3)^\intercal$.}
\label{tab:SMfieldcontent}
\end{table}

In the limit of vanishing Yukawa interactions, the non-universal gauge structure yields the accidental flavor symmetry
\begin{align}
 U(2)^5\equiv U(2)_q\times U(2)_\ell\times U(2)_u\times U(2)_d\times U(2)_e\,.
\end{align}
This is also an approximate symmetry of the SM Yukawa sector~\cite{Barbieri:2011ci,Blankenburg:2012nx,Barbieri:2012uh}, offering a good starting point for the explanation of the observed SM Yukawa hierarchies~\cite{Bordone:2017bld,Bordone:2018nbg,Greljo:2018tuh}. Moreover, the approximate accidental $U(2)^5$ flavor symmetry provides a protection mechanism against the stringent flavor constraints. Indeed, in the absence of fermion mixing terms (see Section~\ref{sec:TC4321}), only third-generation fermions couple to the $U_1$ leptoquark, while $Z^\prime$ and $G^\prime$ couplings to light-generation fermions are suppressed for large $g_4$. As a result, the NP scale can lie around the TeV without conflicting any low-energy or high-$p_T$ bound~\cite{Cornella:2019hct}. Furthermore, this flavor structure is well compatible with the observed NP hints in $B$-meson decays~\cite{Cornella:2019hct,Fuentes-Martin:2019mun}.

\section{Technicolored 4321}\label{sec:TC4321}

We assume that the 4321 gauge group is broken to the SM subgroup \`a la technicolor by confining strong dynamics. Such breaking pattern is minimally realized in $SU(N)_{HC}$ with $4$ hyper-quark flavors that transform in complex and vector-like representations of HC.\footnote{Alternative implementations of a technicolor-like breaking of the 4321 gauge group are discussed in Appendix~\hyperref[app:4321]{B a}.} We denote the hyper-quarks by $\zeta$, and assume for simplicity that they transform in the fundamental of HC. This model has an $SU(4)_L\times SU(4)_R\times U(1)_V$ global symmetry that we partially gauge to reproduce the 4321 gauge sector. More precisely, we identify $SU(4)_R$ with the fully gauged $SU(4)$ and $SU(4)_L$ with the partially gauged $SU(4)'$, which contains $SU(3)^\prime \times U(1)'$ as a subgroup (cf. Figure~\ref{fig:4321breaking}). The $U(1)_X$ symmetry is a combination of $U(1)'$ with another $U(1)$, which together with $SU(2)_L$ belongs to the sector of the theory discussed in Section~\ref{sec:Higgs}. This partial gauging fixes the 4321 representations of the $\zeta$ hyper-quarks (see Table~\ref{tab:TCsector}). 

Similarly to the QCD case, once the HC group becomes strongly coupled, the $\zeta$ hyper-quarks form a condensate
\begin{align}\label{eq:CondensateZeta}
\langle \bar\zeta_L^\alpha\,\zeta_R^\beta\rangle = -\frac{1}{2}\,B_\zeta\, f_\zeta^2\,\delta_{\alpha\beta}\approx -4\pi f_\zeta^3\,\delta_{\alpha\beta}\,,
\end{align}
with $\alpha,\beta=1,2,3,4$, and where $B_\zeta$ and $f_\zeta$ are non-perturbative constants with dimension of energy. The hyper-quark condensate triggers the symmetry breaking
\begin{align}\label{eq:4321TCBreaking}
    SU(4)_L\times SU(4)_R\times U(1)_V\to SU(4)_V\times U(1)_V\,,
\end{align}
where $V$ denotes the diagonal $L+R$.
As a result, also the 4321 gauge symmetry is broken dynamically to the SM gauge group, with $SU(3)_c$ and part of $U(1)_Y$ corresponding to a subgroup of the unbroken $SU(4)_V$, which can be identified with $SU(4)_D$ shown in Figure~\ref{fig:4321breaking}. All the NGBs associated with the chiral symmetry breaking are eaten by the massive gauge bosons. Furthermore, similarly to what happens in technicolor, this dynamical breaking implies $f_U\approx f_{Z^\prime}\approx f_{G^\prime}\approx f_\zeta$ for the gauge boson masses in~\eqref{eq:4321GaugeMasses}. This is a result of the approximate hyper-custodial symmetry $SU(4)_V$. In analogy with the $\rho$ parameter in the SM, we define the quantities
\begin{align}\label{eq:rhoParameters}
\rho_{1,3}&\equiv \frac{M_U^2}{M_{Z^\prime,G^\prime}^2\cos^2\theta_{1,3}}\,,
\end{align}
which are predicted to be 1 in the absence of $SU(4)_V$ breaking sources. Since $\cos\theta_{1,3}$ is completely fixed in terms of SM gauge couplings and $g_4$, we have ${M_U\lesssim M_{Z^\prime}<M_{G^\prime}}$. In particular, in the limit $g_4\gg g_c$ the heavy gauge boson spectrum is quasi-degenerate, while for $g_4\approx g_c$ there is a large splitting between the coloron mass and that of the other two gauge bosons. The gauging of QCD and some of the extended HC interactions (see below) introduce an explicit breaking of the global $SU(4)_V$ symmetry. As a result, the relation among $f_{U,Z^\prime,G^\prime}$ receives loop corrections proportional to these breaking sources. Given the smallness of these corrections, the relation $\rho_1\approx\rho_3\approx1$ is a robust prediction of our setup. The same symmetry breaking pattern and heavy gauge bosons masses are reproduced by the vev of fundamental scalars $\Omega_1$ and $\Omega_3$ as in~\eqref{eq:4321vevs}, satisfying the relation $\omega_1=\omega_3$. However, an important difference with respect to this setup is the absence of scalar radial excitations, i.e. the analogous to the SM Higgs boson in the $\Omega_{1,3}$ fields. This further implies different predictions of the corresponding oblique parameters of the $U_1$, $G^\prime$ and $Z^\prime$ gauge bosons. 

\begin{table}[t]
\begin{center}
\begin{tabular}{|c|c|c|c|c|c|}
\hline
Field & $SU(N)_{\rm HC}$ & $SU(4)$ & $SU(3)'$ & $SU(2)_L$ & $U(1)_X$ \\
\hline
\hline
$\zeta_R$ & $\tiny\yng(1)$ & $\mathbf{4}$ & $\mathbf{1}$ & $\mathbf{1}$ & $0$ \\
$\zeta^q_L\oplus\zeta^\ell_L$ & $\tiny\yng(1)$ & $\mathbf{1}$ & $\mathbf{3}\oplus\mathbf{1}$ & $\mathbf{1}$ & $1/6\oplus-1/2$  \\[1pt]
\hline
\hline
$\chi_L^i$ & $\mathbf{1}$ & $\mathbf{4}$ & $\mathbf{1}$ & $\mathbf{2}$ & $0$ \\ 
$\chi_R^{q\,i}\oplus\chi_R^{\ell\,i}$ & $\mathbf{1}$ & $\mathbf{1}$ & $\mathbf{3}\oplus\mathbf{1}$ & $\mathbf{2}$ & $1/6\oplus-1/2$ \\[1pt]
\hline
\end{tabular}
\end{center}
\caption{Matter content and gauge symmetry representations for the technicolor-like sector. The HC-singlet fermions in the lower blocks, with family index $i=1,\dots,N/2$, are introduced to cancel gauge anomalies.}
\label{tab:TCsector}
\end{table}

Apart from the $\zeta$ hyper-quarks, additional HC-singlet fermions, which we denote by $\chi$, are required to  cancel gauge anomalies. These play a similar role to that of the leptons in the SM. The requirement of anomaly cancellation completely fixes the transformation properties of these fermions under $SU(4)\times SU(3)^\prime$ and their multiplicity in terms of the number of hypercolors $N$, but there is freedom in the choice of transformations under the EW gauge sector. For a specific choice of EW quantum numbers, the fermions $\chi$, which are vector-like under the SM gauge group, can mix with the would-be SM fermions. This allows them to perform two important tasks. First, they induce $U_1$ couplings to the light-generation SM fermions through this mixing. Second, in the presence of Higgs Yukawa couplings with $\chi$ and third-generation SM fermions, the same mixing also generates the 2-3 entries of the CKM matrix. Both of these effects are phenomenologically required, either to explain the $B$ anomalies or to reproduce the structure of the CKM matrix. In other models based on the 4321 gauge group, the introduction of such vector-like fermions is ad hoc. We stress that in our construction, the $\chi$ fermions serve to cancel gauge anomalies and are thus not only phenomenologically motivated but theoretically required, and that, furthermore, their multiplicity is determined by the number of hypercolors $N$. Moreover, in contrast to other 4321 models, the mass of these fermions is not arbitrary and can be connected to the scale of the HC condensate, as we show below. 

The right mixing between $\chi$ and would-be SM fermions to perform the two tasks mentioned above is obtained when we arrange the $\chi$ fermions in $N/2$ families of $SU(2)_L$ doublets (see Table~\ref{tab:TCsector}). With this choice of quantum numbers, a mass-mixing term between the left-handed SM-like families and the $\chi$ fermions is allowed,
\begin{align}\label{eq:ChiLightMassMixing}
\mathcal{L}\supset - M_q\,(\bar q_L\chi_R^q) - M_\ell\,(\bar\ell_L\chi_R^\ell)\,,
\end{align}
with $M_{q,\ell}$ being $2\times N/2$ matrices in flavor space. The minimal phenomenologically viable implementation is obtained for $N=4$, yielding one family of $\chi$ for each SM-like family~\cite{Bordone:2018nbg}. The mass mixing terms induce a coupling between the corresponding SM fields and the $U_1$ leptoquark.\footnote{Note that we could have chosen to identify the $SU(4)$ gauge factor with $SU(4)_L$ instead of with $SU(4)_R$. In that case, no couplings between the SM light generations and the $U_1$ leptoquark would be generated through these mass mixing terms.} As anticipated, this coupling is phenomenologically required for the explanation of the $B$~anomalies. The mixing terms explicitly break the accidental $U(2)^5$ flavor symmetry, unless $M_q^{ij}\propto M_\ell^{ij}\propto\delta_{ij}$ such that the $\chi$ fermions appropriately transform under this symmetry. We assume the existence of an extended hypercolor~(EHC) sector, which generates the following four-fermion interactions
\begin{align}\label{eq:ETHchiMassOP}
\mathcal{L}_{\rm EHC}\supset\frac{1}{\Lambda_{\rm EHC}^2}\;c_\chi^{q,\ell}\,(\bar\chi_L\chi_R^{q,\ell})(\bar\zeta_L^{q,\ell} \zeta_R)\,,
\end{align}
where $\Lambda_{\rm EHC}$ is the EHC scale, and $c_\chi^{q,\ell}$ are $N/2\times N/2$ matrices. This operator induces a technicolor-like mass for the $\chi$ fields after condensation,
\begin{align}\label{eq:ETHchiMass}
M_\chi^{q,\ell}\approx \frac{4\pi f_\zeta^3}{\Lambda_{\rm EHC}^2}\,c_\chi^{q,\ell}\,.
\end{align}
The simultaneous presence of $M_{q,\ell}$ and $M_\chi^{q,\ell}$ yields a collective breaking of the $U(2)^5$ flavor symmetry, irrespective of the form of $M_{q,\ell}$. We take $M_{q,\ell}\ll M_\chi^{q,\ell}$ so that the mass mixing between SM-like and $\chi$ fermions is small, and $U(2)^5$ still remains a good approximate symmetry. Even in this case, one typically requires that the relation $M_q^{ij}\propto\delta_{ij}$ is approximately respected to pass the stringent constraints from $\Delta F=2$ observables~\cite{Cornella:2019hct,DiLuzio:2018zxy,Bordone:2018nbg}. To avoid LHC constraints on the new QCD-colored fermions, we require $M^q_\chi$ to lie at the TeV scale. This in turn implies $\Lambda_{\rm EHC}\approx10~\mathrm{TeV}$ for $\mathcal{O}(1)$ couplings and $f_\zeta\approx2.5~\mathrm{TeV}$ (a value that is motivated by the fit to the $B$ anomalies~\cite{Cornella:2019hct}).

We also introduce the following four-fermion operators
\begin{align}\label{eq:ETH3mixing}
\mathcal{L}_{\rm EHC}\supset\frac{1}{\Lambda_{\rm EHC}^2}\;c_{\psi\chi}^{q,\ell}\,(\bar\psi_L\chi_R^{q,\ell})(\bar\zeta_L^{q,\ell} \zeta_R)\,,
\end{align}
with $c_{\psi\chi}^{q,\ell}$ being $N/2$-dimensional vectors, which induce a mass mixing between $\chi_L$ and $\psi_L$ after HC condenses. This mixing is expected to be sizable if the EHC dynamics generating these operators is the same that also generates the operators in~\eqref{eq:ETHchiMassOP}. As with the mixing between $\chi$ and the SM-like families, this is a welcome feature for the explanation of the $B$ anomalies~\cite{Cornella:2019hct}.

Additional higher dimensional operators could in principle also be generated by the same EHC dynamics. The most relevant are operators of the type $(\bar\psi_L\chi_R)^2$, $(\bar\psi_L\chi_R)(\bar\chi_R\chi_L)$ or $(\bar\chi_L\chi_R)^2$. After mass mixing, these would induce one loop contributions to flavor violating four-fermion processes with two light and two third-generation SM fermions, or with four light-generation SM fermions. The approximate $U(2)^5$ flavor symmetry is enough to prevent them from violating current flavor bounds for the assumed value of $\Lambda_{\rm EHC}$.

\section{The composite Higgs sector}\label{sec:Higgs}

The naturalness problem in the Higgs sector is solved if the Higgs boson is a composite state of strong dynamics confining at a scale $\Lambda$ not far from the TeV scale. The splitting $m_h\ll\Lambda$, required for phenomenological reasons, is achieved if the Higgs arises as a pNGB from the spontaneous breaking of an (approximate) global symmetry of the strong dynamics. We consider the possibility that the same strong dynamics triggering 4321 spontaneous symmetry breaking also produces such composite Higgs boson. 

The simplest implementation of this idea consists in having different HC representations for the hyper-quarks triggering 4321 breaking and for those generating the Higgs. This way, the global symmetry group factorizes, minimizing the number of pNGBs. We focus on the minimal composite Higgs implementation, and discuss other options in Appendix~\hyperref[app:CHM]{B b}. We fix $N=6$, and introduce four chiral fermions $\xi$ in the pseudoreal three-index antisymmetric representation of $SU(6)_{\rm HC}$, $A_3=\boldsymbol{20}$.\footnote{Another interesting realization is obtained for $N=4$ with $5$ hyper-quarks in the real $A_2=\boldsymbol{6}$ representation. This gives a pNGB composite Higgs via the $SU(5)\to SO(5)$ global symmetry breaking, producing more pNGBs than in the minimal model presented here (cf. Appendix~\hyperref[app:CHM]{B b}).
A similar composite Higgs sector is discussed in~\cite{Ferretti:2013kya,Ferretti:2014qta,Golterman:2015zwa,Golterman:2017vdj,Agugliaro:2018vsu} and analyzed on the lattice~\cite{DeGrand:2015lna,DeGrand:2016htl,Ayyar:2017qdf,Ayyar:2018zuk,Ayyar:2018ppa,Cossu:2019hse}.} The transformations of the $\xi$ hyper-quarks under the 4321 gauge group are given in Table~\ref{tab:Higgs}. Note that $N=6$ corresponds to three families of $\chi$ HC singlets (see Table~\ref{tab:TCsector}). Alternatively, we can arrange the HC singlets into two families of $\chi$ and one family of $\tilde\chi$, transforming as $\chi$ except that they are $SU(2)_L$ singlets and their $U(1)_X$ charges are shifted by $\pm \frac{1}{2}$. The latter option allows to modify the couplings of the heavy vector bosons to right-handed third-generation fermions via mass mixing, analogously to the mixing with $\chi$ discussed in Section~\ref{sec:4321}. In the following, we focus on this latter option, i.e. we consider HC singlets in two families of $\chi$ and one family of $\tilde{\chi}$. Note that the extra fermion content yields a loss of $SU(2)_L$ asymptotic freedom. This happens already with the $\chi$ fermions alone, and it is a common behavior in most 4321 models in the literature. With the matter content introduced here, the $SU(2)_L$ Landau pole is at around $10^{14}~\mathrm{GeV}$. It is thus conceivable that this group will unify into a larger group (or that the matter content will split into several groups) at a scale below this Landau pole.

\begin{table}[t]
\begin{center}
\begin{tabular}{|c|c|c|c|c|c|}
\hline
Field & $SU(6)_{\rm HC}$ & $SU(4)$ & $SU(3)'$ & $SU(2)_L$ & $U(1)_X$ \\
\hline
\hline
$\xi_L$ & $\boldsymbol{20}$ & $\mathbf{1}$ & $\mathbf{1}$ & $\mathbf{2}$ & $0$ \\
$\xi_R^+\oplus \xi_R^-$ & $\boldsymbol{20}$ & $\mathbf{1}$ & $\mathbf{1}$ & $\mathbf{1}$ & $1/2\oplus -1/2$ \\[2pt]
\hline
\end{tabular}
\end{center}
\caption{Gauge symmetry representations for the hyper-quarks producing the pNGB Higgs.}
\label{tab:Higgs}
\end{table}

Once we introduce the new hyper-quarks, the composite sector presents an additional $SU(4)_{\rm EW}\times U(1)_A$ global symmetry. The $U(1)_A$ factor corresponds to an anomaly-free combination of the axial symmetries of the $\zeta$ and $\xi$ fields, cf.\ e.g.~\cite{DeGrand:2016pgq}. The $SU(4)_{\rm EW}$ symmetry contains the $SU(2)_L\times SU(2)_R$ symmetry as a subgroup, which is partially gauged to give the $SU(2)_L\times U(1)_X$ factors in $\mathcal{G}_{4321}$. More precisely, $SU(2)_L$ is identified with the one in the 4321 model, and the $U(1)_X$ charge is defined by the combination $X=T_R^3+\sqrt{2/3}\,T_{4^\prime}^{15}$, where $T_R^3$ is the diagonal $SU(2)_R$ generator and $T_{4^\prime}^{15}$ is the corresponding diagonal generator of the $SU(4)^\prime$ symmetry. Contrary to the $\zeta$ hyper-quarks, the new fermions can have  mass terms. These read 
\begin{align}\label{eq:PsiMasses}
\mathcal{L}\supset - M_L\,\bar\xi_L^{i\,c}\,\epsilon_{ij}\,\xi_L^j - M_R\,\bar\xi_R^{i\,c}\,\epsilon_{ij}\,\xi_R^j\,,
\end{align}
where the $c$ superscript denotes charge conjugation and $\epsilon_{ij}$ is the antisymmetric tensor of $SU(2)_{L,R}$. The masses $M_{L,R}$ are taken to be real and positive. This can be done without loss of generality by an appropriate field redefinition. These masses explicitly break the global $SU(4)_{\rm EW}\times U(1)_A$ symmetry. However, they do not break the $SU(2)_L\times SU(2)_R$ subgroup. 

The $\xi$ hyper-quarks also form condensates once HC becomes strongly coupled
\begin{align}\label{eq:CondensateXi}
\braket{\bar\xi_L^{i\,c}\,\xi_L^j}=\braket{\bar\xi_R^{i\,c}\,\xi_R^j}=-\frac{1}{2}\,B_\xi\, f_\xi^2\,\epsilon_{ij}\approx -4\pi f_\xi^3\,\epsilon_{ij}\,,
\end{align}
with $B_\xi$ and $f_\xi$ being non-perturbative constants with dimension of energy, different from $B_\zeta$ and $f_\zeta$ in~\eqref{eq:CondensateZeta}, but expected to be of similar size. This condensate triggers the spontaneous global symmetry breaking\footnote{Fundamental composite Higgs models with the $SU(4)\to Sp(4)$ symmetry breaking have first been discussed in~\cite{Kaplan:1983sm} and more recently in~\cite{Katz:2005au,Gripaios:2009pe,Galloway:2010bp,Barnard:2013zea,Ferretti:2013kya,Cacciapaglia:2014uja,Cacciapaglia:2020kgq}. Lattice studies can be found in~\cite{Lewis:2011zb,Hietanen:2013fya,Hietanen:2014xca,Arthur:2016dir,Bennett:2017kga,Bennett:2019jzz,Bennett:2019cxd}.} 
\begin{align}\label{eq:HiggsGlobal}
SU(4)_{\rm EW}\times U(1)_A\to Sp(4)_{\rm EW}\,,
\end{align}
resulting in $6$ pNGBs: the Higgs doublet, $H$, and two real singlets, $\eta_1$, related to the $U(1)_A$ breaking, and $\eta_5$. The unbroken $Sp(4)\cong SO(5)$ symmetry contains the partially gauged $SO(4)\cong SU(2)_L\times SU(2)_R$ as a subgroup. This global subgroup contains the diagonal $SU(2)_V$ custodial symmetry, which protects the ratio of $W$ and $Z$ masses against corrections from the strongly-coupled dynamics. There is another alignment of the $\xi$ condensate of physical interest: $\braket{\bar\xi_L^i\,\xi_R^j}\propto\delta_{ij}$, analog to the one in~\eqref{eq:CondensateZeta}.
In contrast to the condensate in~\eqref{eq:CondensateXi}, this one breaks the EW symmetry to its $U(1)_{\rm em}$ subgroup, and is the condensate used in minimal technicolor models~\cite{Weinberg:1975gm,Susskind:1978ms,Farhi:1980xs}. As we discuss in Section~\ref{subsec:HiggsPot}, radiative corrections induced by the Yukawa interactions tend to align the vacuum along the technicolor direction. The resulting misalignment from the EW preserving direction gives rise to a successful EW symmetry breaking triggered by the composite Higgs.

\subsection{Yukawa interactions}\label{subsec:Yukawas}

To generate Yukawa couplings between the elementary fermions and the composite Higgs, the two sectors must be coupled. In modern composite Higgs models, this is usually done by introducing mixing terms between composite fermions and the would-be SM fermions. After mixing, the SM states are then \textit{partially composite}, and the required Higgs Yukawas are generated. This approach requires light composite partners for the top. In fundamental composite Higgs models, these partners should correspond to composite baryons, whose natural mass scale is close to the compositeness scale ${\Lambda_{\rm HC}\approx4\pi f_\xi/\sqrt{6}}$, far too heavy to generate enough mixing for the large top mass. Even if one could argue for a large mixing, our setup does not have fermionic baryons in the absence of strongly-coupled scalars~\cite{Sannino:2016sfx}. Alternatively, one can couple the elementary fermion bilinears directly to scalar operators of the strong sector. These couplings can arise from four-fermion operators involving two hyper-quarks and two elementary fermions, analogous to those in Section~\ref{sec:TC4321}. 

To this end, we introduce the following EHC interactions
\begin{align}\label{eq:ETHYukawa}
\mathcal{L}_{\rm EHC}&\supset\frac{1}{\Lambda_{\rm EHC}^2}\left\{\left[y^+_\psi\,(\bar\psi_L\psi_R^+)+y_{ \chi\psi}^+(\bar\chi_L\psi_R^+)\right](\bar\xi_R^+\,\xi_L)\right.\nonumber\\
&\quad\left.+\left[y^-_\psi\,(\bar\psi_L\psi_R^-)+y_{ \chi\psi}^-(\bar\chi_L\psi_R^-)\right](\bar\xi_R^-\,\xi_L)\right\}\,.
\end{align}
Here $y^\pm_\psi$ are numbers and $y_{ \chi\psi}^\pm$ are $2$-dimensional vectors in flavor space. Once the HC group confines, the scalar current of hyper-quarks is interpolated to a composite Higgs field, giving rise to Higgs Yukawa interactions for the would-be third-family SM fermions and the $\chi$. To reproduce the observed top mass with an $\mathcal{O}(1)$ coupling, the EHC scale for the corresponding operator should be of similar size to the one in~\eqref{eq:ETHchiMassOP}, that is $\Lambda_{\rm EHC}\approx10~\mathrm{TeV}$. The smallness of bottom and tau masses compared to the top mass requires either the EHC scale for the corresponding interactions to be larger, or having a large mixing between the $\tilde\chi$ fermions and right-handed bottom and tau. Note that the Yukawa interactions above give $m_b=m_\tau$ in the absence of fermion mixing. Experimentally, one finds
\begin{align}
\left.\frac{m_b-m_\tau}{m_b}\right|_{\mu=2~\mathrm{TeV}}\approx0.2\,,
\end{align}
close to the unification condition. The fermion mixing with $\chi$ and (or) $\tilde\chi$ introduces $SU(4)$-breaking sources that modify the tau-bottom mass relation. These can easily accommodate the (small) mass difference. The 2-3 entries in the CKM matrix are generated via the second and fourth operator in~\eqref{eq:ETHYukawa} through the mixing between $\chi^q$ and the light-generation SM-like quarks. The smallness of this CKM matrix element is naturally explained if this mixing is small, as we assumed in the previous section. Alternatively, one could have a larger mixing, and a (slightly) larger EHC scale for these operators. The approximate $U(2)^5$ flavor symmetry protects the model against large flavor violating contributions from possible four-fermions operators of the form $(\bar\psi_L\psi_R)^2$ and $(\bar\chi_L\psi_R)^2$, keeping them below current flavor bounds. On the other hand, four-fermion operators with only $\xi$ fields would produce a breaking of the global symmetry similar to the one in~\eqref{eq:ETHYukawa}.

Light-generation masses are obtained via the EHC operators
\begin{align}\label{eq:ETHYukawaLight}
\mathcal{L}_{\rm EHC}\supset\frac{1}{\Lambda_{\rm EHC}^{\prime\,2}}&\left[y_u(\bar q_L u_R)(\bar\xi_R^+\,\xi_L)+y_d(\bar q_L d_R)(\bar\xi_R^-\,\xi_L)\right.\nonumber\\
&\left.+\,y_e(\bar \ell_L e_R)(\bar\xi_R^-\,\xi_L)\right]\,,
\end{align}
where $y_{u,d,e}$ are $2\times2$ matrices in flavor space. We assume that there is a large separation of scales between the dynamics generating these operators and those introduced before, namely $\Lambda_{\rm EHC}\ll\Lambda_{\rm EHC}^\prime$. From the charm Yukawa coupling, we estimate this scale to be $\Lambda_{\rm EHC}^\prime\approx100~\mathrm{TeV}$. This assumption,  which is entirely  motivated  by  the  observed  SM  Yukawa  hierarchies, gives a $U(2)$-like protection from other possible four-fermion  operators  involving  only  SM-like  fields. Such protection is enough to pass the stringent bounds from $\Delta F=2$ observables, provided these operators receive a loop suppression compared to the ones in~\eqref{eq:ETHYukawaLight}. This could easily be achieved if the mediators generating the EHC operators are charged under HC. Alternatively, one could obtain this additional suppression if the strong sector is close to a conformal fixed point and the condensate has a sizable anomalous dimension, as in walking technicolor~\cite{Holdom:1981rm}. 

Finally, we could also introduce EHC operators of the form
\begin{align}
\mathcal{L}_{\rm EHC}&\supset\frac{1}{\Lambda_{\rm EHC}^{\prime\prime\,2}}\left[c_q\,(\bar q_L\chi_R^q)+c_\ell\,(\bar\ell_L\chi_R^\ell)\right](\bar\xi_R^{+\,c}\,\xi_R^-)\,,
\end{align}
with $c_{q,\ell}$ being $2\times2$ flavor matrices, and analogously with $(\bar\xi_L^{+\,c}\,\xi_L^-)$. After HC condensation, these would give mass mixing terms as in~\eqref{eq:ChiLightMassMixing}, together with a Yukawa coupling to the pNGB singlets. The smallness of these terms compared to those in~\eqref{eq:ETHchiMassOP} can be explained by requiring $\Lambda_{\rm EHC}\ll\Lambda_{\rm EHC}^{\prime\prime}$. As in the case above, this scale separation provides a sufficient flavor protection for possible operators with two SM-like and two $\chi$ fields. Similarly, operators with $\tilde\chi$ and $\psi_R^\pm$ would introduce mass-mixing terms analogous to~\eqref{eq:ChiLightMassMixing}, which are already present at the level of renormalizable interactions, or four-fermion interactions that are not phenomenologically relevant.

\subsection{The pNGB potential}\label{subsec:HiggsPot}

The compositeness scale sets the masses of most of the composite particles. The pNGBs constitute an exception since their mass is protected by the global symmetry. The potential for these bosons is proportional to the different explicit symmetry breaking terms: the $\xi$ fermion masses, the 4321 gauging, and the EHC four-fermion operators. In this section we discuss the pNGB masses obtained from these breaking terms, and the necessary conditions for EW symmetry breaking.

Like the quark masses in the QCD chiral Lagrangian, the hypercolored fermion masses in~\eqref{eq:PsiMasses} provide an explicit global symmetry breaking and give masses to the pNGBs. Using spurion analysis for the pNGB Lagrangian, we find
\begin{align}
m_H^2&=B_\xi\,(M_L+M_R)\,.
\end{align}
The singlets mix for $M_L\neq M_R$, similarly to what happens with the $\pi^0$ and the $\eta$ in QCD. Their squared-mass matrix in the $(\eta_1\;\,\eta_{5})$ basis reads
\begin{align}
m_\eta^2=B_\xi
\begin{pmatrix}
\tilde q_\xi^{\,2}(M_L+M_R) & \tilde q_\xi\,(M_L-M_R)\\[5pt]
\tilde q_\xi\,(M_L-M_R) & (M_L+M_R)
\end{pmatrix}
\,.
\end{align}
We defined $\tilde q_\xi\equiv 2\,q_\xi f_\xi/f_1$, where $q_\xi=-1/\sqrt{10}$ is the $U(1)_A$ charge of the $\xi$ hyper-fermions and $f_1$ is the decay constant of $\eta_1$, normalized as in~\cite{DeGrand:2016pgq}. The $\xi$ fundamental masses provide the dominant contribution to the pNGB singlet masses. Moreover, as we discuss below, they are also needed to obtain a phenomenologically viable breaking of the EW symmetry.

The explicit breaking of the global symmetry due to the 4321 gauging also yields contributions to the pNGB potential. This is analogous to the gauging of electromagnetism in the QCD chiral Lagrangian, responsible for the mass splitting of pions and kaons. Analogously to the QCD case~\cite{Bijnens:1996kk}, the $\eta_5$ singlet does not receive mass corrections from the gauging, while those to $\eta_1$ are suppressed in the large $N$ limit. Therefore, gauge corrections are only relevant for the Higgs.
They give positive contributions to the pNGB mass squared and hence do not induce a vacuum misalignment~\cite{Witten:1983ut}. 

In order to misalign the vacuum, fermion-loop contributions to the pNGB potential induced by the EHC operators are required. The operators in~\eqref{eq:ETHYukawa} do not
induce a vev for $\eta_5$, but they could do so for $\eta_1$. 
For simplicity, we assume that this is not the case and leave a more general analysis for future work. A vanishing vev of $\eta_1$ would follow automatically, for instance, if the model parameters are chosen such that $\eta_1$ is a CP eigenstate. Hence, to study the potential, we set to zero all fields except the physical Higgs boson, $h$. In this case, the fluctuations around the EW preserving vacuum can be parameterized by $\theta=\theta_{\rm min}+h/f_\xi$.
The Coleman-Weinberg potential~\cite{Coleman:1973jx} for $\theta$ reads
\begin{align}\label{eq:HiggsPot}
V(\theta)&\approx-C_m\,f_\xi^4\,\cos\theta-(C_y-C_g)\,f_\xi^4\,\sin^2\theta\,,
\end{align}
with the following definitions
\begin{align}\label{eq:C_ymg}
\begin{aligned}
C_y&\approx\frac{2\,\Lambda_{\rm HC}^2}{16\pi^2 f_\xi^2}\sum_i c_i\, |\hat y_i|^2\,,\qquad
C_m=\frac{m_H^2}{f_\xi^2}\,,\\
C_g&=\frac{3\,\Lambda_{\rm HC}^2}{32\pi^2 f_\xi^2}\left(\frac{3}{4}\,g_L^2\,c_L +\frac{1}{4}\,g_Y^2\,c_Y\right)\,,
\end{aligned}
\end{align}
where the index $i$ spans the EHC interactions in~\eqref{eq:ETHYukawa}, $\hat y_i\approx y_i\,4\pi f_\xi^2/\Lambda_{\rm EHC}^2$ are the Higgs Yukawa couplings, and $c_{L,Y,i}$ are non-perturbative coefficients expected to be of $\mathcal{O}(1)$ and positive. The angle $\theta_{\rm min}$ parameterizes the orientation of the true vacuum between the EW preserving and the technicolor vacuum. The EW symmetry is unbroken when $\theta_{\rm min}=0$, while for $\theta_{\rm min}=\pi/2$ we obtain a technicolor breaking. As can be seen from~\eqref{eq:HiggsPot}, the $\xi$ fundamental masses and gauge radiative contributions tend to align the vacuum along the EW preserving direction. The Yukawa contributions tend to align it along the technicolor direction, provided the non-perturbative coefficients are indeed positive. A non-zero $\theta_{\rm min}$ is obtained when $C_m<2(C_y-C_g)$. Minimizing the potential gives the EW symmetry breaking condition 
\begin{align}\label{eq:HiggsTuning}
\frac{v^2}{f_\xi^2}\equiv\sin^2\theta_{\rm min}=1-\frac{C_m^2}{4\,(C_y-C_g)^2}\,,
\end{align}
with $v\approx246~\mathrm{GeV}$ corresponding to the SM Higgs vev. To achieve the desired $\sin\theta_{\rm min}$ value, this condition has to be tuned by appropriately choosing the HC masses and Yukawa couplings.  Note that the fit to the $B$ anomalies suggests that $f_\zeta\in[2.5,4]~\mathrm{TeV}$~\cite{Cornella:2019hct}. Hence, since we expect $f_\xi\approx f_\zeta$, this translates into a $1\%$ tuning for the lowest value of $f_\zeta$. The mass of the Higgs boson is readily obtained from the potential and reads
\begin{align}\label{eq:HiggsMass}
m_h^2=2\left(C_y-C_g\right)v^2\,.
\end{align}
Note that this expression is independent of $f_\xi$ once the Higgs vev $v$ is fixed, since $\Lambda_{\rm HC}\propto f_\xi$ in \eqref{eq:C_ymg}. In the limit where $C_y$ is saturated by the top Yukawa contribution, we can rewrite the expression above as
\begin{align}
m_h^2\approx\frac{4}{3}\,c_\psi^+\,m_t^2-\frac{3}{2}\,c_L\,m_W^2-\frac{1}{2}\,c_Y\,(m_Z^2-m_W^2)\,,
\end{align}
where we took $\Lambda_{\rm HC}\approx4\pi f_\xi/\sqrt{6}$. The non-perturbative coefficients $c_i$ are not free parameters and can be determined from the HC dynamics. At present, no determination of these coefficients is available, but naive dimensional analysis suggests that no further tuning seems necessary for the Higgs mass, once the tuning in~\eqref{eq:HiggsTuning} is achieved.

The main constraints on the value of $\sin\theta_{\rm min}$ are obtained from EW precision tests, Higgs coupling modifications, and the modification to the $Zb_L\bar b_L$ coupling. Constraints from EW precision tests and Higgs coupling modifications for this class of models are discussed in~\cite{Cacciapaglia:2020kgq}, yielding $\sin\theta_{\rm min}\lesssim0.2\,$. A somewhat stronger bound is usually obtained from the modification to the $Zb_L\bar b_L$ coupling, whose experimental limit is at the per mille level. This constraint does not apply to our setup, unless the EHC sector contains vector operators producing a direct coupling between the vector resonances and the SM fermions.  In any case, all these constraints are satisfied if $f_\xi$ is around the TeV.

\section{Conclusions}\label{sec:Conclusions}

Models of low-scale partial unification have recently regained interest due to the anomalies in $B$-physics data. Indeed, one of the most compelling solutions to this puzzle consists in extending the SM gauge symmetry to the so-called 4321 gauge group, allowing for natural low-scale unification of third-family quarks and leptons. 

Interestingly, the NP scale and flavor structure suggested by these anomalies hint at a possible connection with the solution of the hierarchy problem. Moreover, as we argue in this letter, there is a striking similarity between the 4321 and EW gauge sectors. This has taken us to consider the possibility of dynamically breaking both symmetries by hyper-quark condensates of the same strong dynamics, which we denoted as HC. The simplest way to realize this idea requires two sets of hyper-quarks transforming under different representations of the HC group. This way, each set is responsible for either 4321 or EW symmetry breaking. The minimal implementation for each of these sets is given by the $\zeta$ and $\xi$ fermions in Tables~\ref{tab:TCsector} and~\ref{tab:Higgs}. The similarity between the 4321 and the EW sector is also manifest in the hyper-quark quantum numbers for these minimal implementations. However, a crucial difference between the two is that, while the $\zeta$ belong to a complex HC representation, the $\xi$ belong to a pseudoreal one. This difference is instrumental in delaying a technicolor-like breaking of the EW symmetry, thus explaining the mass gap between 4321 and EW massive gauge bosons, and producing a pNGB composite Higgs.

Apart from introducing a composite Higgs and providing a dynamical mechanism for the 4321 symmetry breaking, the model presented here has also several other appealing features not found in most 4321 models discussed in the literature. Most importantly, it offers a theoretical motivation for the $\chi$ fermions, vector-like under the SM gauge group. The existence of these fermions and the requirement of having their mass close to the 4321 breaking scale, as needed for the phenomenological viability of these models, are ad-hoc features in most realizations. In our model, the $\chi$ fermions are theoretically required to cancel HC anomalies, analogously to the leptons in the SM, and their mass is connected to the 4321 breaking scale through the EHC operators in~\eqref{eq:ETHchiMassOP}. The phenomenology of 4321 models has been discussed in many places and a recent analysis can be found in~\cite{Cornella:2019hct}. We leave a detailed phenomenological discussion to future work, but note that we have ensured that our construction satisfies all the requirements to reproduce a phenomenology similar to that in~\cite{Cornella:2019hct}. A major difference between our model and other 4321 models discussed in the literature, including the one in~\cite{Cornella:2019hct}, is the prediction of mass relations for the heavy gauge bosons (see~\eqref{eq:rhoParameters}), analogous to the $\rho$ parameter in the EW sector. Current low-energy and high-$p_T$ data is consistent with this prediction. However, this is a smoking gun signature that could be tested in the near future.

There are several directions that require future investigation. One of the main challenges in the construction presented here consists in finding a well-motivated description of the dynamics responsible for the EHC operators. The chiral structure of the operators in~\eqref{eq:ETHchiMassOP} suggests that this might be in the form of bosonic EHC, analogous to bosonic technicolor~\cite{Samuel:1990dq}. However, we note that the protection from the approximate $U(2)^5$ flavor symmetry, inherited from the 4321 gauge structure, effectively eliminates the flavor problem common to many of these solutions. Another interesting avenue is the possibility of having composite Dark Matter. The composite spectrum contains a SM-singlet baryon of spin $1$, consisting of a $(\boldsymbol{6}\,\boldsymbol{6}\,\boldsymbol{6}\,\boldsymbol{20})$ hyper-quark bound state, that could potentially play the role of a Dark Matter candidate. If this is the lightest baryonic resonance, its stability is guaranteed provided the $U(1)_V$ symmetry in~\eqref{eq:4321TCBreaking} remains unbroken, analogously to the proton in QCD. These \textit{chimera baryons}, composed of fermions in two different HC representations, are indeed expected to be the lightest baryonic resonances.\footnote{This has been show on the lattice for a strongly-coupled $SU(4)$ with fermions in the fundamental and two-index antisymmetric representation~\cite{Ayyar:2018zuk}.} The lightness of the SM-singlet spin-$1$ baryon compared to other chimera baryons could be explained by mass corrections due to the QCD (and EW) gauging, since all chimera baryons of spin $0$ are colored. Ultimately, a lattice study of the baryonic spectrum is required to confirm this possibility.

The LHCb and Belle~II experiments will give a definite answer to the nature of the $B$-physics anomalies in the next few years. While we believe that this model stands out as an interesting theoretical framework on its own, if they are confirmed as genuine NP effects, it could provide one of the most motivated explanations for the $B$~anomalies.

\section*{Acknowledgements}

We thank Gino Isidori, David Marzocca, Francesco Sannino and, in particular,  Admir Greljo for carefully reading the manuscript and for useful comments. The work of JFM has received funding from the Swiss National Science Foundation (SNF) under contract 200021-175940, by the European Research Council (ERC) under the European Union's Horizon 2020 research and innovation programme, grant agreement 833280 (FLAY), and by the Generalitat Valenciana under contract SEJI/2018/033. PS is grateful for the support and hospitality of the Pauli Center for Theoretical Studies and the University of Zurich.

\appendix

\section{SM fermion embedding in 4321 models} \label{app:SMfermionEmbedding}

The SM fermion content can be arranged in different representations under the extended gauge symmetry, yielding different 4321 model implementations. If we restrict to fundamental representations, each SM family admits three possibilities:
\begin{enumerate}[label=\roman*)]
    \item \textit{Pati-Salam-like representation}, where both chiralities of quarks and leptons are unified into 4-plets of $SU(4)$. In this case, the corresponding fermions are singlets of $SU(3)^\prime$, and transform as in the Pati-Salam model~\cite{Pati:1974yy} under the $SU(4)\times SU(2)_L\times U(1)_X$ subgroup. 
    \item \textit{SM-like representation}, where quarks and leptons are singlets of $SU(4)$, and transform as in the SM under the $SU(3)^\prime\times SU(2)_L\times U(1)_X$ subgroup. In this case, no direct couplings to the $U_1$ leptoquark are present.
    \item \textit{Mixed representation}, where one of the chiralities of quarks and leptons are unified in a 4-plet, while the other remains SM-like. This option requires additional fermions to cancel gauge anomalies. These could be SM fermions from another family, or new fermions that either acquire a mass which is larger than the EW scale or which are charged under a new strong interaction. Moreover, in this case the SM Higgs needs to be embedded in a bi-fundamental representation under $SU(4)\times SU(3)^\prime$ or the Yukawa couplings have to be provided through the mixing with additional vector-like fermions.
\end{enumerate}

Several 4321 implementations have been recently considered in the literature in connection with the $B$ anomalies. In the models in~\cite{DiLuzio:2017vat,DiLuzio:2018zxy}, the three families of would-be SM fermions (when neglecting the mixing with other fermions) are arranged in SM-like representations. Couplings to the $U_1$ leptoquark are induced via mass mixing with heavy vector-like fermions that are charged under the $SU(4)$ subgroup. An example of mixed representation can be found in a low-energy limit of the model in~\cite{Fornal:2018dqn}. In this model, all three SM-like families are arranged in mixed representations, and additional matter is introduced to render the model anomaly free. In contrast to the previous realizations, in the models in~\cite{Cornella:2019hct,Greljo:2018tuh}, corresponding to the low-energy limit of the Pati-Salam cubed model~\cite{Bordone:2017bld}, the would-be SM families transform differently under the extended gauge symmetry. More precisely, the third family is arranged in the Pati-Salam-like representation, while the other two families are arranged in SM-like representations (see Table~\ref{tab:SMfieldcontent}). As in the models in~\cite{DiLuzio:2017vat,DiLuzio:2018zxy}, one can introduce mass mixing with heavy vector-like fermions to induce $U_1$ leptoquark interactions with the light families and (or) Yukawa interactions among third and light families.

\section{Model variations}\label{app:Variations}

In the main part of this letter, we introduce a minimal model that realizes a technicolor-like breaking of the 4321 symmetry, contains a composite Higgs, and features a well-motivated fermion sector, which renders the model free from gauge anomalies and preserves asymptotic freedom of the new strong interaction. There are various variations of the minimal setup that might be interesting to explore in future work. We summarize several possibilities in this appendix.

A main idea of this letter is to use a single strongly-coupled sector to both break the 4321 gauge group and to generate a pNGB Higgs. The breaking of the 4321 gauge group is achieved by embedding it into the global symmetries of the strong sector in such a way that it is spontaneously broken to the SM gauge group once the global symmetries are broken by a fermion condensate. To generate a viable pNGB Higgs, the breaking of the global symmetries further has to yield a pNGB with the quantum numbers of the Higgs doublet and leave the custodial $SU(2)_L\times SU(2)_R$ symmetry unbroken. In general, there are several possibilities for such global symmetry breaking in the chiral limit of strongly-coupled gauge theories. The different patterns of global symmetry breaking in the cases of massless fermions transforming in complex, real, or pseudoreal representations of the strongly-coupled gauge group are given by~\cite{Peskin:1980gc,Preskill:1980mz,Kosower:1984aw}:
\begin{enumerate}[label=\roman*)]
    \item \textit{Complex and vector-like representation} with $2 N_f$ Weyl or $N_f$ Dirac fermions:
    \begin{equation}
        SU(N_f)_L\times SU(N_f)_R\to SU(N_f)_V\,,
    \end{equation}
    with $N_f^2-1$ NGBs.
    \item \textit{Real representation} with $N_f$ Weyl or Majorana fermions:
    \begin{equation}
        SU(N_f)\to SO(N_f)\,,
    \end{equation}
    with $(N_f-1)(N_f+2)/2$ NGBs.
    \item \textit{Pseudoreal representation} with $2 N_f$ Weyl or $N_f$ Dirac fermions:
    \begin{equation}
        SU(2 N_f)\to Sp(2 N_f)\,,
    \end{equation}
    with $(N_f-1)(2 N_f+1)$ NGBs.
\end{enumerate}
The scenarios yielding the smallest number of NGBs are those in which fermions transforming in two different representations of the strong gauge group are each responsible for either the 4321 symmetry breaking or the generation of the composite Higgs.

\paragraph{4321 symmetry breaking.}\label{app:4321}
For the breaking of the 4321 symmetry, giving masses to the $U_1$, $Z'$ and $G'$, one can use either a complex, real, or pseudoreal representation. Some of the resulting NGBs are would-be NGBs that act as longitudinal polarizations of the heavy gauge bosons, while the remaining ones are pNGBs that receive masses of several TeV from gauge boson loops. The minimal constructions are:
\begin{enumerate}[label=\roman*)]
    \item \textit{Complex representation:} $SU(4)\times SU(4)\to SU(4)$
    
    The gauged $SU(4)\times SU(3)'$ and part of $U(1)_X$ are embedded in the initial $SU(4)\times SU(4)$, while $SU(3)_c$ and part of $U(1)_Y$ are embedded in the unbroken diagonal $SU(4)$. The 15 NGBs transform under $SU(3)_c\times U(1)_Y$ as $\boldsymbol{8}_0$, $\boldsymbol{3}_{+2/3}$, $\boldsymbol{\bar{3}}_{-2/3}$, and $\boldsymbol{1}_0$, and they all correspond to would-be NGBs.
    
    \item \textit{Real representation:} $SU(8)\to SO(8)$
    
    The gauged $SU(4)\times SU(3)'$ and part of $U(1)_X$ are embedded in an $SU(4)\times SU(4)$ subgroup of $SU(8)$, while $SU(3)_c$ and part of $U(1)_Y$ are embedded in an $SU(4)$ subgroup of the unbroken $SO(8)$. The 35 Nambu-Goldstone bosons transform under $SU(3)_c\times U(1)_Y$ as $\boldsymbol{8}_0$, $\boldsymbol{3}_{+2/3}$, $\boldsymbol{\bar{3}}_{-2/3}$, $\boldsymbol{1}_0$, which are would-be NGBs, and $\boldsymbol{1}_{\pm1}$, $\boldsymbol{3}_{-1/3}$, $\boldsymbol{\bar{3}}_{+1/3}$, $\boldsymbol{6}_{+1/3}$, and $\boldsymbol{\bar{6}}_{-1/3}$, which are pNGBs.
    
    \item \textit{Pseudoreal representation:} $SU(8)\to Sp(8)$
    
    The gauged $SU(4)\times SU(3)'$ and part of $U(1)_X$ are embedded in an $SU(4)\times SU(4)$ subgroup of $SU(8)$, while $SU(3)_c$ and part of $U(1)_Y$ are embedded in an $SU(4)$ subgroup of the unbroken $Sp(8)$. The 27 NGBs transform under $SU(3)_c\times U(1)_Y$ as $\boldsymbol{8}_0$, $\boldsymbol{3}_{+2/3}$, $\boldsymbol{\bar{3}}_{-2/3}$, $\boldsymbol{1}_0$, which are would-be NGBs, as well as $\boldsymbol{3}_{\pm1/3}$ and $\boldsymbol{\bar{3}}_{\pm1/3}$, which are pNGBs.
\end{enumerate}
A different way of breaking the 4321 symmetry without invoking scalar fields could be realized in terms of a tumbling gauge group~\cite{Raby:1979my} that breaks itself. A 4321 gauge group that breaks itself to the SM has been described in~\cite{Martin:1992mj} in the context of topcolor and technicolor models.

\paragraph{Composite Higgs.}\label{app:CHM}

For the composite Higgs sector, the requirements of generating a NGB with the quantum numbers of the Higgs doublet and preserving the custodial $SU(2)_L\times SU(2)_R$ symmetry leads to the following minimal constructions using a complex, real, or pseudoreal representation:
\begin{enumerate}[label=\roman*)]
    \item \textit{Complex representation:} $SU(4)\times SU(4)\to SU(4)$
    
    The 15 NGBs transform as $\boldsymbol{3}_0$, $2\times\boldsymbol{2}_{\pm\frac{1}{2}}$, $\boldsymbol{1}_\pm$ and $2\times\boldsymbol{1}_0$ under $SU(2)_L\times U(1)_Y$.
    
    \item \textit{Real representation:} $SU(5)\to SO(5)$
    
    The 14 NGBs transform as $\boldsymbol{2}_{\pm\frac{1}{2}}$, $\boldsymbol{3}_0$, $\boldsymbol{3}_\pm$, and $\boldsymbol{1}_0$ under $SU(2)_L\times U(1)_Y$.
    
    \item \textit{Pseudoreal representation:} $SU(4)\to Sp(4)$
    
    The 5 NGBs transform as $\boldsymbol{2}_{\pm\frac{1}{2}}$ and $\boldsymbol{1}_0$ under $SU(2)_L\times U(1)_Y$.
\end{enumerate}
For each of these cases, minimal composite Higgs models have been constructed and analyzed in detail: see for instance~\cite{Ma:2015gra,Agugliaro:2019wtf} for
i),~\cite{Georgi:1984af,Dugan:1984hq,Ferretti:2013kya,Ferretti:2014qta,Golterman:2015zwa,Golterman:2017vdj,Agugliaro:2018vsu,Agugliaro:2019wtf} for ii), and~\cite{Kaplan:1983sm,Katz:2005au,Gripaios:2009pe,Galloway:2010bp,Barnard:2013zea,Ferretti:2013kya,Cacciapaglia:2014uja,Cacciapaglia:2020kgq,Redi:2012ha,Alanne:2014kea,Gertov:2015xma,Serra:2015xfa,Cai:2015bss,Galloway:2016fuo,Agugliaro:2016clv,Ferretti:2016upr,Niehoff:2016zso,Cacciapaglia:2017cdi,Sannino:2017utc} for iii).

\paragraph{4321 symmetry breaking with a composite Higgs.}

It is possible to use fermions transforming under a single representation of the strong gauge group for both breaking the 4321 symmetry and generating a pNGB Higgs. In this case, the number of pNGBs increases compared to the cases in which two different representations are employed. However, such models have a simpler UV structure. The minimal constructions are: 
\begin{enumerate}[label=\roman*)]
    \item \textit{Complex representation:} $SU(8)\times SU(8)\to SU(8)$
    
    featuring 63 NGBs:
    \begin{itemize}
        \item 15 like in the composite Higgs scenario where ${SU(4)\times SU(4)\to SU(4)}$,
        \item 15 like in the 4321 breaking scenario where ${SU(4)\times SU(4)\to SU(4)}$,
        \item 1 real singlet plus 16 complex scalars with the same quantum numbers as one full generation of SM fermions (plus an $SU(2)_L$-singlet neutrino).
    \end{itemize}

    \item \textit{Real representation:} $SU(11)\to SO(11)$
    
    featuring 90 NGBs:
    \begin{itemize}
        \item 14 like in the composite Higgs scenario where ${SU(5)\to SO(5)}$,
        \item 35 like in the 4321 breaking scenario where ${SU(8)\to SO(8)}$,
        \item 1 real singlet plus 20 complex scalars with SM quantum numbers $(\boldsymbol{3}, \boldsymbol{2})_{2/3}$, $(\boldsymbol{3}, \boldsymbol{2})_{-1/3}$, $(\boldsymbol{3}, \boldsymbol{1})_{1/6}$, $(\boldsymbol{1}, \boldsymbol{2})_{0}$, $(\boldsymbol{1}, \boldsymbol{2})_{-1}$, $(\boldsymbol{1}, \boldsymbol{1})_{-1/2}$.
    \end{itemize}
     
    \item \textit{Pseudoreal representation:} $SU(12)\to Sp(12)$
    
    featuring 65 NGBs:
    \begin{itemize}
        \item 5 like in the composite Higgs scenario where ${SU(4)\to Sp(4)}$,
        \item 27 like in the 4321 breaking scenario where ${SU(8)\to Sp(8)}$,
        \item 1 real singlet plus 16 complex scalars with the same quantum numbers as one full generation of SM fermions (plus a $SU(2)_L$-singlet neutrino).
    \end{itemize}
\end{enumerate}

\newpage
\bibliographystyle{JHEP}
\bibliography{references}
\end{document}